# High Speed and High Efficiency Travelling Wave Single-Photon Detectors Embedded in Nanophotonic Circuits

W. H. P. Pernice[1*], C. Schuck[1], O. Minaeva[2], M. Li[1], G.N. Goltsman[3], A.V. Sergienko[2], H. X. Tang[1,+]

[1]*Department of Electrical Engineering, Yale University, New Haven, CT 06511, USA*

[2]*Department of Electrical & Computer Engineering, Boston University, Boston, MA 02215, USA*

[3]*Department of Physics, Moscow State Pedagogical University, Moscow 119992, Russia*

[+] Correspondence and requests for materials should be addressed to H.X.T. (email: hong.tang@yale.edu)

**Ultrafast, high quantum efficiency single photon detectors are among the most sought-after elements in modern quantum optics and quantum communication. High photon detection efficiency is essential for scalable measurement-based quantum computation[1-3], quantum key distribution[4], and loophole-free Bell experiments[5]. However, imperfect modal matching and finite photon absorption rates have usually limited the maximum attainable detection efficiency of single photon detectors[6]. Here we demonstrate a superconducting nanowire detector[7] atop nanophotonic waveguides which allows us to drastically increase the absorption length for incoming photons. When operating the detectors close to the critical current we achieve high on-chip single photon detection efficiency up to 91% at telecom wavelengths, with uncertainty dictated by the variation of the waveguide photon flux. We also observe remarkably low dark count rates without significant compromise of detection efficiency. Furthermore, our detectors are fully embedded in a scalable silicon photonic circuit[8-10] and provide ultrashort timing jitter of 18ps. Exploiting this high temporal resolution we demonstrate ballistic photon transport in silicon ring resonators. The direct implementation of such a detector with high quantum efficiency, high detection speed and low jitter time on chip overcomes a major barrier in integrated quantum photonics.**

---

[*] Current address: Karlsruhe Institute of Technology (KIT), Institute of Nanotechnology, 76344 Eggenstein-Leopoldshafen, Germany

Single photon detectors (SPDs) with high detection efficiency play a crucial role in modern quantum optics and information science that utilize photonic qubits as information carriers[11]. The availability of fast photon-counting detectors with close-to-unity efficiency is a prerequisite for measurement-based quantum computation and will significantly enhance the throughput of optical quantum communication systems[12]. On a more fundamental level, highly efficient SPDs will also allow for loophole-free Bell experiments[5] and guarantee that protocols based on quantum nonlocality outperform their classical counterparts[13]. The need for high efficiency photon detectors has recently become ever more pressing as quantum computation and communication applications come into reach of current technology, which is evident from recent progress in developing integrated photonic circuitry[14,15]. However, universal and scalable quantum computation is possible with single photons and linear optics only if efficient qubit measurements can be performed. At present low photon detection efficiencies limit the success probability of quantum gate operations and therefore the scalability of quantum optical circuits. Both the generation of highly entangled graph states for one-way quantum computation[16] and the measurement-induced multi-qubit operations in teleportation-based schemes[1] crucially rely on large numbers of high-efficiency single photon detectors, which to-date are still realized off-chip. Therefore a scalable and efficient detector architecture is highly desirable to move beyond few qubits operations.

Besides efficiency and scalability, high temporal resolution is another key feature required in quantum optics and quantum information processing. The manipulation and processing of photon qubits at GHz rates is becoming a necessity for quantum computers and quantum communication systems to pace up with their classical counterparts[17]. Single photon detectors combining high efficiency with high speed will lead to improved performance of active feed forward[18], which is an essential element of cluster state quantum computing and error correction schemes, as well as entanglement manipulation and quantum communication over longer distances with higher bit rates[19]. Similarly, a major challenge in integrated quantum photonics lies in reliably discriminating alternative optical paths for generating time-bin or path entanglement where picosecond time resolution is desired.

Current state-of-the-art single photon detectors are still far away from reaching such a combination of high speed and efficiency. Silicon avalanche photo-detectors (APDs), that are widely used for visible light detection, have, at best, quantum efficiency of 65%, accompanied by significant dead time (> 65ns)[20] and rather high timing jitter[21]. While single photon measurement in the visible spectrum is important for a wide range of applications, efficient detectors in the near infrared (NIR) and telecom wavelength region (TWR) are necessary to employ the modern optical fiber infrastructure for quantum communication purposes. InGaAs single photon detectors can count TWR photons but usually have only moderate quantum efficiency (~10-25%)[22]. Furthermore, such detectors are plagued by high noise (dark counts) and must run in gated mode. So far the best contender in terms of quantum efficiency is the transition edge sensor (TES)[23]. Unfortunately, these devices have limited speed and must be operated at millikelvin temperatures.

Here we present an integrated single photon detector that operates at 1550nm wavelength with high detection efficiency, high speed and low timing jitter. We employ superconducting nano-wires to achieve single-photon resolution at more easily obtainable liquid Helium temperature. Using a travelling wave design[24] provides nearly full absorption of incoming photons and allows for high counting rates. Efficient coupling between propagating optical waves and the superconducting thin film allows us to drastically reduce the detector length and therefore leads to a small foot-print device that can be efficiently combined with established integrated optical circuits. Despite variation in the coupling efficiency from device to device, we obtain reasonable repeatability in term of on-chip quantum efficiency, which can be further improved by better control in coupler fabrication. Our detectors are realized on commercial silicon-on-insulator substrates and thus allow for realizing densely integrated photonic circuitry. With on-chip optical ring resonators, we are able to resolve the delay of ballistic photon transport with unprecedented temporal resolution, and therefore showcase the outstanding jitter performance of our detectors.

Traditional superconducting nanowire SPDs (SSPDs) absorb incoming photons under normal incidence. Therefore the detector absorption length is determined by the thickness of the superconducting thin film. This implies that only a fraction of the photon flux is absorbed because the film is normally only several nanometers thick. Resonant cavities have been employed to enhance the absorption by virtually prolonging the interaction length[25]. However, this approach inevitably reduces the detector optical bandwidth and the obtainable quantum efficiency is eventually limited by the quality of the deposited end mirrors and the coupling conditions to the cavity.

Here, in contrast, we employ the travelling wave geometry shown schematically in Fig.1a) in order to maximize the waveguide-detector interaction length and thus obtain strong photon absorption. Photons propagate along silicon waveguides in an on-chip photonic network towards the detector region. The aspect ratio of the waveguides are designed to only support the fundamental quasi-TE mode at 1550nm and prevents waveguiding of TM-polarized modes. In the detection area the silicon layer is covered with a superconducting Niobium Nitride (NbN) thin film of 3.5nm thickness, which couples evanescently to the guided optical mode. To achieve hot-spot formation upon photon absorption, the NbN thin film is patterned into two narrow wires, having a wire width of 100nm (70nm and 85nm in another design) and a wire spacing of 120nm. Light from an external laser source is coupled into the chip using focusing grating couplers (Fig. 1b), pink triangles), guided along low-loss photonic waveguides, and further split into two paths: one path leads towards the detector and the other path is directed to a calibration port. After passing through the detector, any residual light is collected into an additional output port (labeled residual light port in Fig.1b)). Thus, for the purpose of detector calibration, the photonic circuit allows us to estimate absorption losses from the NbN wires. In the scanning electron micrograph (Fig. 1b, inset), the NbN structures are discernable as the light regions, located on top of the silicon photonic circuitry, as also highlighted in the second zoom-in SEM inset. In addition to the optical ports, metal RF-contact pads are used to extract the detector signal from the chip, as shown in the upper half of Fig.1b).

The absorption properties of the detector are first analyzed numerically with three-dimensional finite-difference time-domain (FDTD) simulations as shown in Fig.1c). The fundamental optical mode of a 750x110nm$^2$ silicon waveguide is launched from the left towards the NbN detector region (light structure in the center of the image). Fig. 1d displays the supported single TE optical mode. In the presence of the NbN thin film the evanescent tail of the guided optical mode is coupled strongly to the superconducting wires. In this case the optical fields are more tightly confined to the metallic regions, as shown in the inset of Fig.1e), with significant field enhancement at the NbN sides. This coupling leads to strong absorption of the incoming wave, as presented in the log-plot of the modal intensity in Fig.1c). From the optical simulation we obtain 10.1±0.2dB attenuation within a 10μm long waveguide with NbN wires on top, or 1dB attenuation per μm of detector length. The FDTD simulations also reveal no appreciable reflection (<-28dB) from the onset of the meander structure. This is a result of the near-adiabatic transition from the bare silicon waveguide to the NbN covered waveguide section, where the ultrathin NbN layer compared to the underlying silicon layer gives an area ratio of 1:118.

The simulated absorption characteristics are verified experimentally by measuring photon absorption for detectors of varying length. We characterize the optical attenuation of the device shown in Fig. 1b) using a low-temperature cryostat setup configured with a RF probe and a fiber array (see also supplementary materials). A photonic chip containing 60 devices is mounted on a movable three-axis stage (Attocube Systems) that is hanging under a 1K pot inside a Helium-4 Cryostat. After device cool down, we use a tunable laser source (New Focus 6428) to scan the transmission properties of the optical circuits with continuous-wave light (CW). The light at the control port and the residual light port are detected simultaneously with two external low noise photoreceivers. In Fig.1f) we show the measured absorption spectra for a range of devices with total detector length varying from 5μm to 40μm. As shown in the SEM image in the inset of Fig.1b), each detector comprises two wires in series; therefore the length of the NbN wires on top of the waveguide is roughly twice the length of the detector. In order to obtain the data traces, the input laser is operated in swept wavelength mode and the transmission through the device is continuously recorded. The absorption spectra are measured over the bandwidth of the grating couplers from 1520 nm to 1555nm. An almost flat absorption profile (Fig.1f)) shows that the detector is wavelength insensitive over the measured coupler bandwidth. Small variations in the transmission profile are due to the dispersion and loss properties that are commonly found in nanophotonic waveguides. The noise fluctuations in the lowest traces are a result of the photodetector's noise floor. The attenuation increases exponentially with increasing detector length, as also predicted from the numerical simulations in Fig.1c). The absorption of the detector film can be extracted from the attenuation-length dependence as shown in Fig.1g), revealing an attenuation of 0.98±0.10dB per micrometer detector length. The measured results are in good agreement with the simulated absorption rate (blue triangles in Fig.1g)). From the linear fit of the attenuation we find that a detector with a length of 20μm provides more than 97% absorption of the incoming light.

The performance of our single-photon detectors is assessed in terms of detection efficiency, dark count rate, and timing jitter[6,21]. In order to obtain reproducible results and to know the statistical spread of the device performance we analyze fabricated chips with multiple devices. Before analyzing the detector itself, the measurement setup is calibrated in terms of transmission and device deviation (see supplementary materials). To determine the optical power propagating through the on-chip waveguides, the variation of the coupling efficiency of the grating input and output couplers is assessed. We employ a layout with designated calibration devices positioned in the measuring area. The calibration coupler devices are used to gather the statistical spread of the coupling loss for a fiber-to-chip spacing of 100μm which is achieved after establishing electrical contact with the chip through the RF probes (see supplementary materials). From the measured devices we obtain a coupling loss at the point of maximum transmission of -13.1±0.7dB. We also determine the spread of the average transmission loss, which amounts to -16.4±0.6dB. The average transmission is only important for ultrafast jitter measurements where picosecond laser pulses are employed (the optical bandwidth is 2.9nm for a 1.2ps pulse). Despite of the coupler variations from chip to chip, the power extracted from the two output ports on the same device are identical within 0.2dB (5%) in the absence of detectors.

We analyze the on-chip detection efficiency (DE) by measuring the rate of the detected photons propagating inside the waveguide. We fabricate detectors with varying length in order to evaluate the dependence of the on-chip DE on the detector length. The detector counting rates are measured from 1.7K to 4.2K. The integrated circuit is first optically excited using a CW laser source tunable over the telecom C-band (for devices with high dark count rates near $I_c$ we utilize modulated CW light to avoid interference of dark counts with the true counts, see supplementary materials). The optical power fed into the photonic circuit is measured with a calibrated power meter prior to adding controllable attenuation. We fixed the input optical power at 100μW (-10.15dBm). Subsequently, calibrated optical attenuation of 72.6dB is added using the cascaded attenuators. Details of the attenuator calibration are provided in the supplementary materials.

Taking into account the input loss occurring at the input grating coupler as well as the loss due to the on-chip 50/50 splitter, the resulting total photon flux travelling along the waveguide leading to the on-chip detector is estimated to be ~1 million photons/second. At 2K, the critical current $I_c$ is measured to be 28.2μA for typical 100nm-wide devices. The detectors are current-biased with a low-noise battery powered current source close to the critical current. Two cascaded high-bandwidth electrical amplifiers are used to raise the electrical signal for use with a PicoHarp300 Time-Correlated Single Photon Counting (TCSPC) system (by PicoQuant).

Fig.2 illustrates the dependence of the on-chip DE as a function of detector length and bias current, measured at 2K. The detection efficiency in the figure refers to the probability of photon absorption within the on-chip circuit and has therefore been corrected for the insertion loss, which is obtained in real time from the calibration port. After measuring the overall loss through the nanophotonic device the on-chip optical power is estimated from the calibrated input optical power after the attenuators using the measured coupling efficiency. Before the transmission measurement, the actual attenuation of the input power is calibrated at higher optical input powers as well as at single photon levels (see supplementary materials). The detection efficiency is obtained by monitoring the count rate of the on-chip SSPDs using the PicoHarp. The count rate is monitored over extended periods of time in order to correct for counting errors during the measurement. The bias current is scanned between 50% and 99% of the critical current value for several waveguide detector devices, as shown for the upper traces in Fig.2a). We show measured results for detectors with detector lengths between 20μm and 40μm. Furthermore, in addition to the length we also vary the meander width between 70nm and 100nm. The best on-chip detection efficiency out of all the examined devices on the chip is obtained for a 20μm long device with a meander width of 100nm (labeled as device III). In this case we obtain maximum on-chip DE up to 91% at a bias current of 99% $I_c$, which is close to the expected value for the designed optical absorption of 20dB. We can repeatability obtain high DE from detectors with other widths and lengths. For instance, device I (70nm wide, 40um long, orange symbol) and detector IV (100nm wide, 30um long, green symbols) show a maximum DE of 88% and 87%, respectively. Also included in Fig 2a is a detector with less ideal performance in terms of quantum efficiency (detector II, 85nm wide, 30um long, grey triangles). On the measured chip we do not find apparent improved DE for detectors with increased length, which is most likely due to a higher probability of fabrication imperfections for longer wires. The uncertainty in the determination of the detection efficiency mostly results from a worst-case estimate of the spread of on-chip optical power. We estimate an overall 1-sigma uncertainty in the waveguide optical power of 24% due to the coupler variation (14.9%) and the power splitter (4.0%) as well as the measurement error of the count rate using the PicoHarp300 (5.5%). For comparison we also present the detection efficiency measured for an alternative, more conventional meander wire detector design (device V, purple markers in Fig.2a)) where the detector wires are laid out in a traditional meander structure (see Supplementary Materials) with a total meander length of 600µm. In this case we encounter increased scattering loss and thus reduced detection efficiency of 3% at the highest bias current.

In the travelling wave detector geometry the saturation behavior of the detection efficiency close to the critical current is consistent with previously reported results for traditional SSPDs with NbN wires of similar dimensions at a target wavelength of 1550nm. We observe a monotonic increase of the detection efficiency *DE* when the bias current is increased. From the measured results we find that detectors with narrower meanders show a wider saturation plateau with increasing biasing current (detector I), whereas the wider devices barely show a saturation behavior (detector III, IV)[26,27]. The observed behavior reflects the dependence of the internal quantum efficiency (QE) on the bias current, which then relates to the detection efficiency as *DE=AE×QE*, where AE is the absorption efficiency of the detector. Because the use of the travelling wave geometry allows us to strongly increase the photon coupling between waveguide and NbN nanowire, the total detector length is significantly shorter than in traditional detectors which absorb photons under normal incidence. As a result, the number of incisions occurring over the length of the detector as a result of fabrication imperfections is also reduced. This is reflected by the relatively high critical current measured in our detectors compared to previously reported results[27]. The combination with high absorption efficiency as demonstrated in Fig.1f) provides therefore the ingredients for the observed high on-chip detection efficiency. However, the overall system efficiency of the device is reduced by the coupling loss occurring at the input grating coupler.

In addition to the detection efficiency we measure the dark count rate as a function of biasing current as presented in Fig.2b). The dark count rate shown is obtained for the detector II which has the best performance in terms of dark count rates among all the detectors we sampled. Data shown are results for base temperatures between 1.7K and 4K. At the lowest temperature we find for a bias current of 99% $I_c$ a maximum dark count rate of 50Hz. Higher numbers of dark counts occur at elevated temperatures, consistent with previously reported SSPDs[27]. However, we note that included in the unwanted detection events are also counts due to unshielded ambient background light such as possible leakage through the fiber cladding.

We summarize the measurement results for the detectors shown in Fig.2 in table 1. We compare the detector performance in terms of detection efficiency, dark count rate and the resulting noise equivalent power (NEP), which is defined as $NEP = \frac{h\nu\sqrt{2R_d}}{DE}$, where $h\nu$ is the energy of photon and $R_d$ is the dark count rate. For a bias current of 92% $I_c$ we find a best NEP of $1.1\times10^{-19}$ W/Hz$^{1/2}$ for the 85nm wide, 30μm long detector.

| Length (μm) | Width (nm) | Detection efficiency (%) | Dark count rate (at 99% $I_c$, Hz) | Best NEP (W/Hz$^{1/2}$) |
|---|---|---|---|---|
| 40 | 70 | 88 | 453 | $2.2\times10^{-19}$ @ $0.86 I_c$ |
| 30 | 85 | 53 | 50 | $1.1\times10^{-19}$ @ $0.92 I_c$ |
| 30 | 100 | 87 | 6274 | $4.3\times10^{-19}$ @ $0.83 I_c$ |
| 20 | 100 | 91 | 5886 | $2.5\times10^{-19}$ @ $0.82 I_c$ |

The electrical pulse profiles for four 100nm-wide detectors of different lengths are shown in Fig.3a). In order not to distort the pulse profile we employ high-bandwidth electrical amplifiers with a 3dB bandwidth of 15GHz. In Fig.3a) we show pulse traces resulting from 64 averages of individual pulses. An exemplary single shot trace is shown in the inset of Fig.3a). After an initial fast rise time, as shown on the right side of Fig.3a), the decay time extracted from these profiles increases linearly with detector length as shown in Fig.3b), which is expected due to the increase in kinetic inductance[28]. Due to the smaller kinetic inductance of shorter detectors, the relaxation time of the detector is significantly reduced. For the shortest detector of 10μm length we measure a decay time of 455ps and a FWHM of 505ps. For the longest detector the decay time increases to 1392ps, paired with a FWHM of 1226ps. From the measured decay time we extract the kinetic inductance of the fabricated NbN wires. Using a 50Ω load resistance $R_L$ given by the impedance of the RF transmission line for the device readout, the kinetic inductance $L_k$ is given as $\tau_d = L_k/R_L$ [28]. Normalized to the wire length we obtain a value of 0.91nH/μm, which is consistent with literature results. By using short detector geometries we are therefore able to work in a sub-nanosecond pulse regime. Sub-nanosecond pulse width implies that detection rates in excess of 2GHz may be achieved with our detectors. Because of the high critical current as a result of low defect densities after nano-fabrication latching phenomena are not observed during our measurements.

We next use the detector III to determine the timing jitter of the signal from the NbN wires in order to evaluate the detector performance for optical buffering and time-domain multiplexing. We excite the optical circuit with a sub-picosecond pulsed laser source (1.2ps pulse width), tunable over the telecom C-Band. We determine the timing jitter of the detectors using both a high-speed sampling oscilloscope (Agilent Infiniium with 20GHz sampling rate) and the PicoHarp300 Time-Correlated Single Photon Counting (TCSPC) system. We first determine the intrinsic instrument jitter by employing an electrical self-referencing method (splitting the same electrical pulse in two using a T-connector) and extract the jitter from the Gaussian fit to the measured data. For the sampling oscilloscope we obtain intrinsic jitter of less than 1ps, while the PicoHarp300 yields significantly higher instrument jitter of $\tau_{ph}$ = 19ps. The sampling oscilloscope is then employed in start-stop configuration. The attenuated laser output is split with a fiber 50/50 splitter before being fed into the on-chip devices. The light from one arm of the splitter is fed into a high-speed photodetector with 20GHz bandwidth to provide a stable trigger (start) signal. The light from the second arm is fed into the on-chip detector devices and their electrical output is used as a "stop" signal for the jitter analysis. The oscilloscope is run in histogram mode with a time-window size of 200ps, providing high temporal resolution at a sampling rate of 20GHz. The oscilloscope provides an interpolation function in between sampling intervals and allows for true picosecond histogram resolution. The oscilloscope used during the measurements provides 6GHz bandwidth at an input impedance of 50Ω, which provides thermal noise of 70.4μV rms at 300K. From the pulse front shown in the right panel of Fig.3a) we measure a maximum slope of 8ps/µV (before the electrical amplifier). Therefore the instrument limited jitter resolution is predicted to be 20.7ps at FWHM (further details are provided in the supplementary materials). The measured jitter value of the device is then the jitter between consecutive pulses and is shown in Fig.3c) by the blue squares. Fitting the data with a Gaussian function yields a jitter value of 18.4ps. The measured jitter value is close to the Johnson noise limited jitter. For comparison we also show the jitter measurement obtained with the PicoHarp300 (green circles in Fig.3c)). In this case fitting the measured time-resolved coincidence counting rate with a Gaussian function yields a jitter value of $\tau_m$=50ps. However, the real detector jitter is shadowed by the PicoHarp's instrument timing jitter. Therefore the estimated SSPD jitter extracted from the PicoHarp300 amounts to $\tau_{SSPD} = \sqrt{\tau_m^2 - \tau_{ph}^2}$ =46ps[28]. Contributions to the rather high jitter value result from variation in the discriminator value of the PicoHarp300 as well as the relatively high noise value of the fused broadband inverting amplifiers used by the PicoHarp. Furthermore, the PicoHarp300 cannot be used in TTTR mode in our case due to the high repletion rate of the pulsed-laser source, which also contributes to increased jitter values.

To further verify that the jitter signal is not limited by the bandwidth of the electrical amplifiers, we measure the jitter in dependence of the analog bandwidth of the amplifier set as shown in Fig.3d). Even for low bandwidth of 1.4GHz, the oscilloscope yields a reduced jitter value of 30ps. Increasing the amplifier bandwidth above 10GHz leads to a converged jitter estimate of 18.4ps. Beyond 10GHz, the electrical bandwidth is limited by our cryogenic cables and the detector's impedance mismatch to 50Ω circuits.

The high quantum efficiency and fast detector response of our detector enable time-domain multiplexing in integrated photonic circuits. This opens a way for high-fidelity manipulation of multiple optical qubits on chip. In order to demonstrate the applicability of the SSPDs to fast on-chip single-photon measurements we examine variable photon-delay from a micro-ring resonator. The device used in the experiments is shown in Fig.4a). Two sets of input grating couplers are used to provide measurement capabilities of the micro-ring in both through and drop port configuration. The optical output from one of the drop lines is split with an on-chip 50/50 splitter and fed into a grating output port and an integrated SSPD (shown in the inset of Fig.4a)). The input waveguides are coupled to a ring resonator of 5.8mm in length. The optical resonator is laid out in a meander form with a radius of 25µm at the bends in order to reduce the covered chip area. The bending loss induced at the turning points is small and does not severely compromise the optical quality factor because we are employing silicon waveguides with a high refractive index contrast. The width of the waveguides is fixed at 750nm, which provides measured propagation loss of 4.3dB/cm. Because the final devices comprise a waveguide length below 1mm (excluding the ring resonators), the propagation loss due to surface roughness is not significant for device performance.

The ring resonator can be used in the overcoupled or undercoupled regime by varying the gap between the coupling waveguide and the ring. When the gap is small, the resonator is strongly coupled to the feeding waveguide and thus a significant portion of the input light is transferred into the ring. The light circulating within the ring is, in turn, coupled out efficiently as well. Thus the circulating intensity drops quickly over time, as illustrated in the schematic in Fig.4b). This strong coupling reduces optical quality factors because the cavity is overloaded. We measure the transmission spectrum in the though port with a tunable laser source in order to assess the quality factor of the ring. Results for an overcoupled device with an input gap of 100nm are shown by the purple curve in Fig.4b). Because of the large circumference of the ring resonator the free spectral range (FSR) is small, leading to dense transmission dips at the optical resonances in the spectrum. Fitting the dips with a Lorentzian function yields a quality factor of 14,000 in the overcoupled case (orange curve in Fig.4b)). When the coupling gap $g$ is increased, less light is coupled into and out of the ring resonator. Therefore, light circulating inside the ring decays slower and produces elongated pulse trains when the ring is excited with a pulsed laser source (Fig.4c)). We measure improved optical quality factors around 24,000, as shown in Fig.4c) with the blue markers (the red line denotes the Lorentzian fit to the resonance curve) for a device with increased coupling gap of 200nm.

We then analyze the ring parameters in the time-domain with our low jitter single photon detectors. The optical circuit is excited with attenuated picosecond laser pulses and photon detection events are registered with the on-chip SSPD. The detector is biased far enough from the critical current (86% $I/I_c$) to yield good detection efficiency (~15%) at low dark count rates (<1Hz). The traces are recorded in the drop port (through port traces are also recorded, see Supplementary Materials). For the overcoupled ring resonator we measure the time-domain trace shown in Fig.4d). The photons circulating inside the ring resonator quickly decay from the cavity, so that only 2 peaks are discernible in the linear plot of the arrival time histogram. A third peak is barely visible above the detector count background in the log-plot shown in the inset of Fig.4d). We can obtain the decay time of the ring resonator from the position and height of the peak amplitudes, which amounts to 19.3ps in an exponential fit to the data. Converted into the spectral domain, the decay time corresponds to an optical quality factor of 11,900 which is in good agreement with the measured spectral value. The decay from the cavity is slowed down when we consider an undercoupled device with a coupling gap of 200nm. In this case we are able to observe four consecutive pulse fronts in the time-domain trace. Fitting peak positions with an exponential function reveals a decay time of 37.1ps corresponding to a spectral width of 67pm or equivalently an optical quality factor of 22,900. The positions of the pulse peaks are separated by a delay time of 72.7ps, which is determined by the length of the ring resonator and the group index of the waveguide profile. The group index value measured from the spectrum is 3.58, close to the simulated value of 3.6. The corresponding delay time of 69ps is likewise in good agreement with the measured value.

The demonstrated performance of our fast and high-efficiency superconducting single-photon detector opens a road for integrating tens and hundreds of such devices on a single chip. It will allow for the realization of densely packed integrated photonic circuits for quantum information processing and quantum communication applications. The unique combination of near-unity detection efficiency with increased detector bandwidth and time resolution makes the device invaluable for high-fidelity manipulation and evaluation of quantum states of light on a chip, in both spatial and temporal domains. The use of silicon for device manufacture offers additional advantages in comparison with integrated photonic circuits realized on silica-based substrates because of its large refractive index contrast. Therefore the overall size of quantum photonic circuits can be drastically reduced and combined with other electronic components that are already developed on silicon substrates. While the detectors demonstrated here were analyzed in the telecom band, the detection wavelength range can in principle be extended towards the band-edge limited transmission window of silicon down to 1100nm with adjusted grating couplers. Furthermore, by moving to an alternative material system such as Silicon Nitride we anticipate that our detectors can also be used at visible wavelengths.

We also expect exciting applications by exploiting integrated SSPDs for photon buffering using silicon based photonic circuits. With 4.3dB/cm loss waveguides, we were already able to resolve 4 round trips of photons in a 5.8mm ring. Recent progress in CMOS level nanofabrication has enabled fabrication of ultra-low loss waveguides with propagation loss on the order of 0.1dB/m. Although there are technical challenges to interface such waveguides with our detectors, in principle, it is possible to delay photons for many round trips without experiencing loss in rings of similar dimensions as the devices demonstrated here. Such a low loss delay line also brings photon number resolving detectors within reach, which ideally complement recently demonstrated transition edge detectors[29] and nanowire detectors on GaAs waveguides[30], yet with the benefit of much higher quantum efficiency and speed.

To address the ultimate desires of the quantum communication community the detection speed of our travelling wave detectors can be increased further into the tens of GHz by reducing the length of the NbN covered waveguide, at the cost of somewhat lower detection efficiency. This will eventually allow for bringing quantum cryptography close to the speed of conventional telecommunication equipment.

**Methods Summary**

Our detectors and integrated optical components are realized on commercially available silicon-on-insulator wafers with a buried oxide thickness of 3μm and a top silicon layer of 220nm thickness. The silicon top layer is thinned down to 110nm by oxidation and subsequent wet etch in buffered Oxide etch (BOE). Niobium nitride thin films of 3.5nm thickness are deposited by dc reactive magnetron sputtering in an Ar and $N_2$ atmosphere. The NbN thin film was deposited directly on cleaned SOI substrates without the use of an intermediate buffer layer. The maximum values of the critical film parameters ($T_c$, and $j_c$) are reached at the discharge current value of 300 mA, a partial $N_2$ pressure of $1.7 \times 10^{-4}$ mbar and a substrate temperature of 850°C. The Ar pressure proved to have no substantial effect on the film deposition rate or the film composition.

Thus the pressure level was set to a level allowing for maintaining a stable discharge ($4.5\times10^{-3}$ mbar). At these conditions the deposition rate, defined as the ratio of the film thickness to the deposition time, was 0.5 nm/s. The as-deposited films were characterized by the surface resistance $R_S$ = 500Ω/sq, critical temperature $T_c$ = 10 to 11K, superconducting transition width $\Delta T_c$ ~ 0.3 K, and critical current density $j_c$ = 6 to $7\times10^6$ A/cm$^2$ measured at 4.2K.

Optical photo-lithography using a double-layer lift-off resist (Shipley 1805 + LOR5A) is utilized to define contact pads and alignment marks. E-Beam evaporation of 5nm Cr and 200nm Au is performed and the contact pads are formed by lift-off in NMP. A first Electron-beam lithography step is carried out on an EBPG 5000+ 100kV system using HSQ e-Beam resist to define the detector structures. A timed etching step in ICP RIE using CF$_4$ chemistry is performed to etch through the NbN layer. Subsequently, a second e-Beam lithography is employed to define optical circuitry. The sample is then etched by ICP RIE in Cl$_2$ plasma to define the waveguiding structures in the top silicon film.

Coherent pump light is launched into the chip from a swept wavelength diode laser (New Focus 6428) combined with two optical attenuators (Tektronix OA5002), which provide up to 60dB of optical signal reduction each. The transmitted light from the device is recorded with a low-noise, high frequency photodetector (New Focus 1554-B) for wavelengths around 1550nm. Optical pulses are generated using a lithium niobate high-speed electro-optical modulator (Lucent 2623NA) combined with a pulse generator (HP8133a). For the ring-down measurements a Pritel fiber laser with a pulse width < 1ps and a period of 12.5ns is employed. The central pulse wavelength can be tuned over the telecom C-Band from 1532nm to 1565nm, which is used to match the input wavelength to the central wavelength of the on-chip grating couplers. The SSPDs are current-biased with a battery powered current source (LakeShore101) and a bias-T. Two electrical amplifiers with varying analog bandwidth (by RF-Bay and Picosecond Labs) are used to elevate the electrical signal by up to 47dB. For use with the PicoHarp300 an inverting 15GHz amplifier (Picosecond Labs 5828) is used. The time-domain signal is recorded on an oscilloscope (Agilent Infiniium 54855A).

**Acknowledgement**

This work was supported by a fellowship program from Packard Foundation. W.H.P. Pernice would like to thank the Alexander-von-Humboldt foundation for providing a postdoctoral fellowship and acknowledges support by DFG grant PE1832-1. The authors wish to thank Dr. Mike Rooks, Michael Power, James Agresta and Christopher Tillinghast for assistance in device fabrication and B.M. Voronov for deposition of the NbN films.

**Figure 1:**

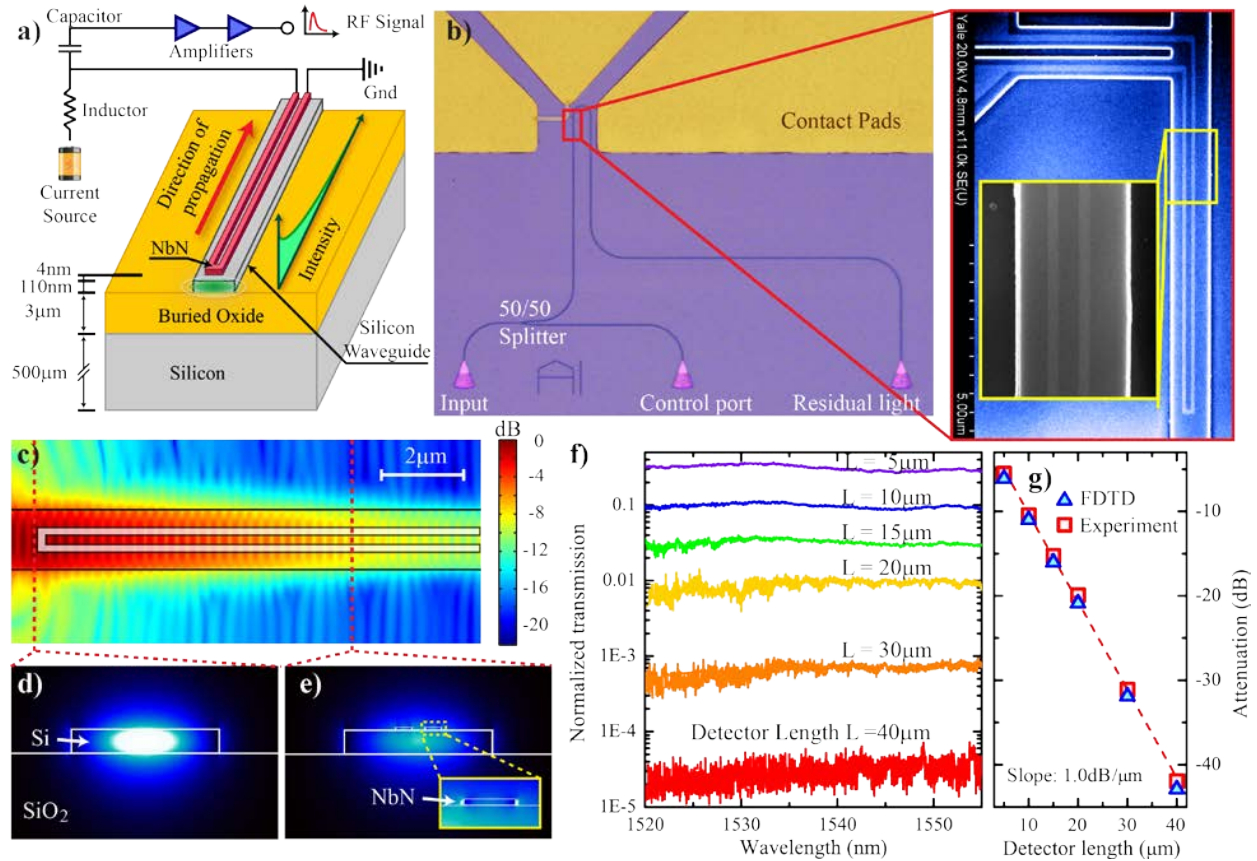

**Figure 1 Design of the travelling wave SSPD.** a) Principle of the travelling wave SSPD: a sub-wavelength absorbing NbN nanowire is patterned atop a silicon waveguide to detect single photons; b) Optical micrograph of a fabricated device showing the optical input circuitry, RF contact pads and the SSPD. The control and residual ports are used for calibration purposes. Inset: zoom into the detector region with an SEM image showing the detector regime. A low density of incisions is maintained due to the short detector length. c) Finite-difference time-domain simulation of the absorption characteristics of the detector. The propagating intensity is displayed in a logarithmic scale. d) The simulated field distribution of the optical mode in the silicon portion of the waveguide; e) The modal pattern with NbN wires on top. The inset shows the optical field concentration around the NbN wires; f) measured absorption spectra of detectors with lengths varying from 5μm to 40 μm in dependence of wavelength. g) The detector attenuation in dependence of length, measured as the ratio of power between the residual port and the control port, exhibits a slope of 0.98dB/μm.

**Figure 2:**

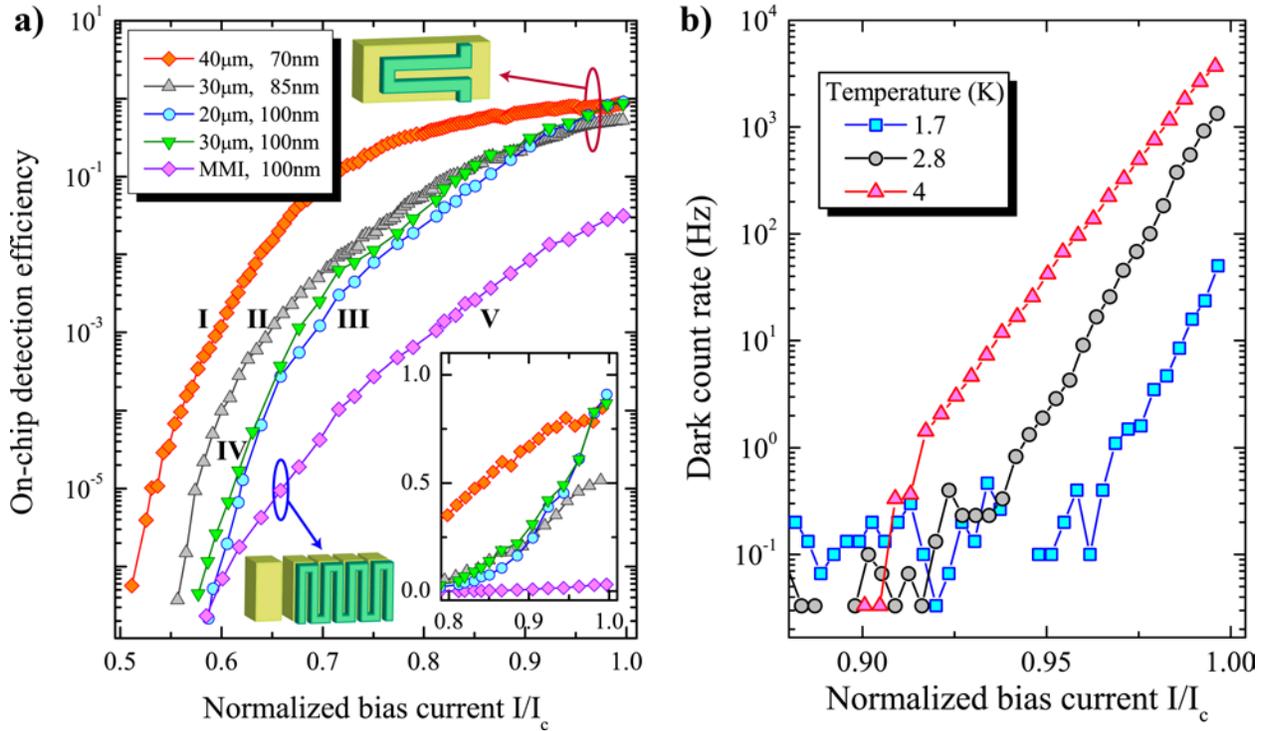

**Figure 2. SSPD on-chip detection performance.** a) Shown is the on-chip detection efficiency of fabricated devices as a function of detector length and normalized biasing current, after correction for coupler and 50/50 splitter loss. The measured devices have meander widths between 70nm and 100nm. The best on-chip detection efficiency of 91% is obtained for a 20μm long, 100nm wide detector (detector III) at 99% of the critical current. For comparison a meander-type detector (MMI) is also measured with a peak detection efficiency of 3%. Inset: A linear plot of the high-biasing current region, showing the monotonic increase of the detection-efficiency towards $I_c$. b) The detector dark count rate measured as a function of biasing current for the grey curve in a) (detector II). Shown are data for three different temperatures between 1.7K and 4K. At 99% $I_c$ the minimum count rate is 50Hz.

**Figure 3:**

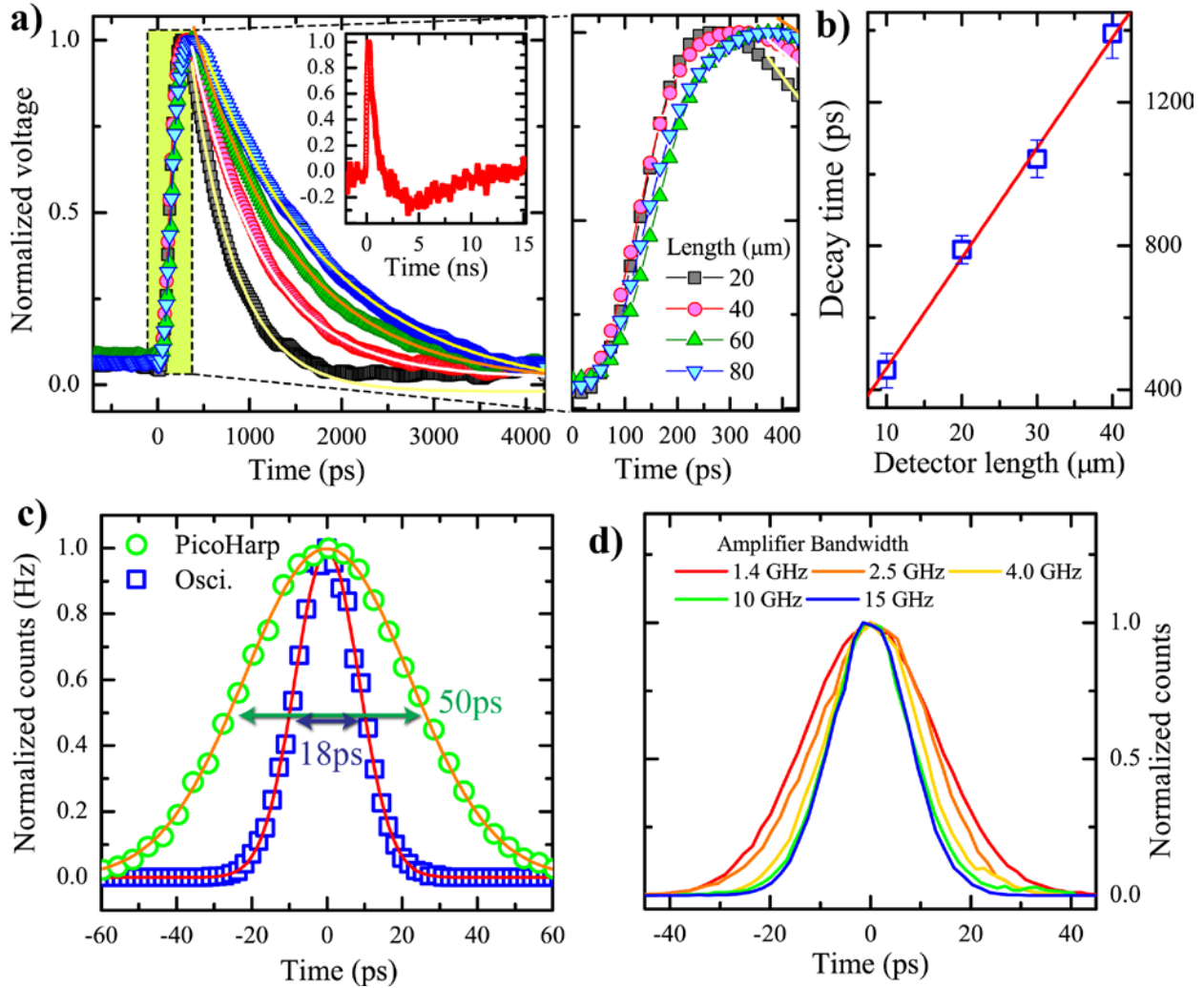

**Figure 3. Single-photon detector characterization in time domain.** a) Measurement of the pulse shape for SSPDs with lengths varying from 10μm to 40μm after 64 averages. Discrete symbols indicate experimental data, the solid lines are exponential fits to the decay of the detector. A decay time of 455ps is obtained for the shortest detector. Inset: Single shot measurement curve. The right side shows the initial slope of the pulse in dependence of detector length. b) The extracted detector decay time as a function of the detector length. The decay time increases linearly with increasing detector length due to increased kinetic inductance. c) The timing jitter measured for a representative detector. Green symbols denote the jitter measured with the PicoHarp300, with an intrinsic instrument jitter of 19ps. Blue symbols denote results obtained with a high-speed oscilloscope with an instrument timing jitter of less than 1ps. d) The timing jitter measured in dependence of amplifier bandwidth. Results converge towards 18.4ps for an amplifier analog bandwidth of 10GHz or more.

**Figure 4:**

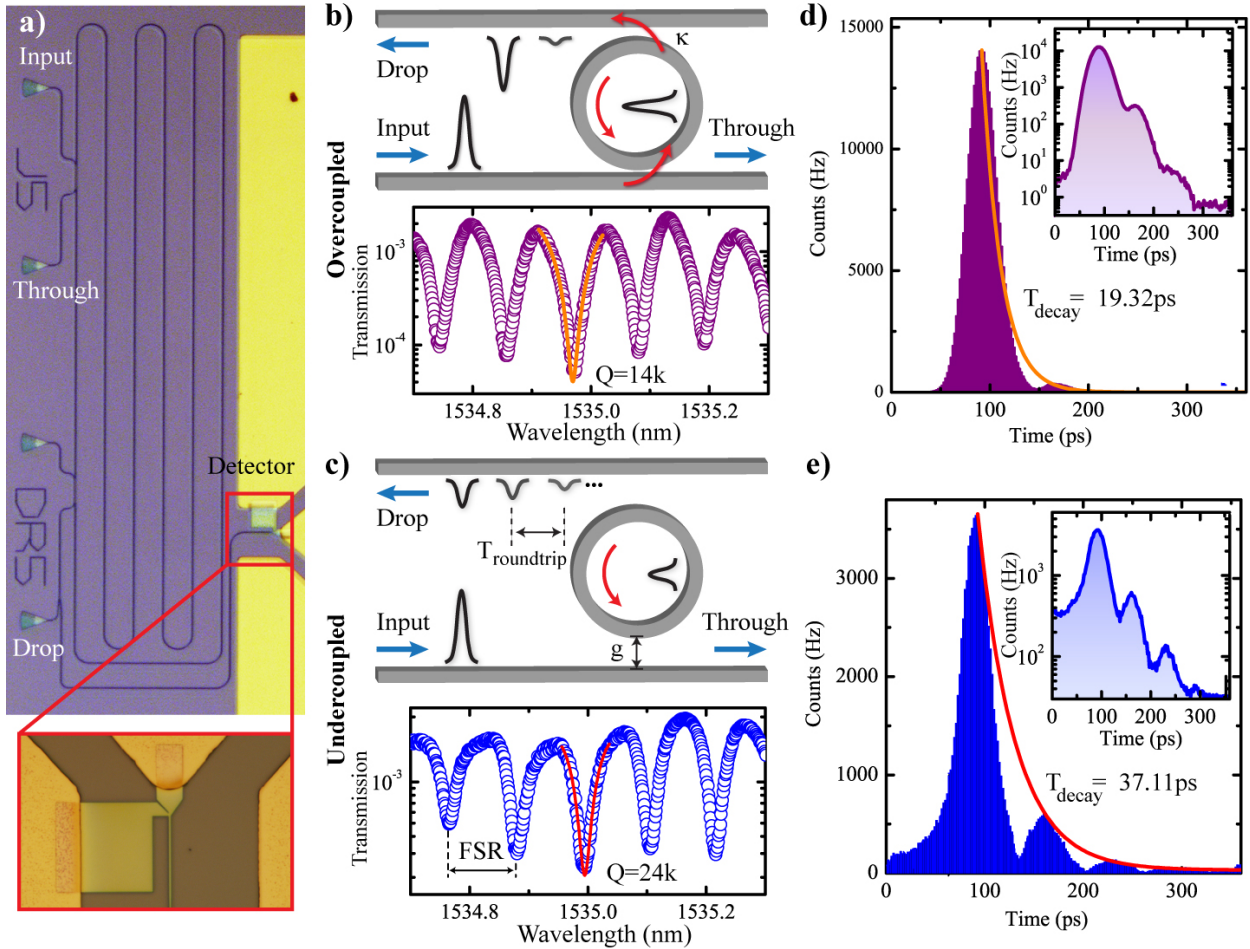

**Figure 4. Time-domain ring-down measurements.** a) An integrated photonic circuit with input grating couplers, a long micro-ring resonator and integrated SSPD. Inset: zoom into the detector region. b) Schematic of the time-domain response of an over-coupled ring resonator. Measured results in the spectral domain reveal an optical quality factor of 14,000. c) Schematic of the time-domain response of a weakly coupled ring resonator in the drop port. The measured optical Q in the frequency domain is 24,000. d) The measured time-domain response for the overcoupled ring resonator. The length of the ring introduces a round-trip delay of 72.7ps, while the circulating pulse amplitude decays with a time constant of 19.3ps when travelling along the ring. e) The measured time-response of the undercoupled ring resonator. The device shows a decay time of 37.1ps. Due to the slower decay four consecutive pulses are discernible in the log-plot (inset).

**References:**
1       Knill, E., Laflamme, R. & Milburn, G. J. A scheme for efficient quantum computation with linear optics. *Nature* **409**, 46-52 (2001).
2       O'Brien, J. L. Optical quantum computing. *Science* **318**, 1567-1570, (2007).
3       Briegel, H. J., Browne, D. E., Dur, W., Raussendorf, R. & Van den Nest, M. Measurement-based quantum computation. *Nat Phys* **5**, 19-26 (2009).
4       Scarani, V. *et al.* The security of practical quantum key distribution. *Rev Mod Phys* **81**, 1301-1350 (2009).
5       Eberhard, P. H. Background Level and Counter Efficiencies Required for a Loophole-Free Einstein-Podolsky-Rosen Experiment. *Phys Rev A* **47**, R747-R750 (1993).
6       Hadfield, R. H. Single-photon detectors for optical quantum information applications. *Nat Photonics* **3**, 696-705 (2009).
7       Sobolewski, R. *et al.* Picosecond superconducting single-photon optical detector. *Appl Phys Lett* **79**, 705-707 (2001).
8       Jalali, B. & Fathpour, S. Silicon photonics. *J Lightwave Technol* **24**, 4600-4615 (2006).
9       Lipson, M., Almeida, V. R., Barrios, C. A. & Panepucci, R. R. All-optical control of light on a silicon chip. *Nature* **431**, 1081-1084 (2004).
10      Vlasov, Y., Green, W. M. J. & Xia, F. High-throughput silicon nanophotonic wavelength-insensitive switch for on-chip optical networks. *Nat Photonics* **2**, 242-246, (2008).
11      Kok, P. *et al.* Linear optical quantum computing with photonic qubits. *Rev Mod Phys* **79**, 135-174 (2007).
12      Varnava, M., Browne, D. E. & Rudolph, T. How good must single photon sources and detectors be for efficient linear optical quantum computation? *Phys Rev Lett* **100**, 060502 (2008).
13      Acin, A. *et al.* Device-independent security of quantum cryptography against collective attacks. *Phys Rev Lett* **98**, 230501 (2007).
14      O'Brien, J. L., Matthews, J. C. F., Politi, A. & Stefanov, A. Manipulation of multiphoton entanglement in waveguide quantum circuits. *Nat Photonics* **3**, 346-350 (2009).
15      Politi, A., Cryan, M. J., Rarity, J. G., Yu, S. & O'Brien, J. L. Silica-on-silicon waveguide quantum circuits. *Science* **320**, 646-649 (2008).
16      Raussendorf, R. & Briegel, H. J. A one-way quantum computer. *Phys Rev Lett* **86**, 5188-5191 (2001).
17      Kim, D., Carter, S. G., Greilich, A., Bracker, A. S. & Gammon, D. Ultrafast optical control of entanglement between two quantum-dot spins. *Nat Phys* **7**, 223-229 (2011).
18      Prevedel, R. *et al.* High-speed linear optics quantum computing using active feed-forward. *Nature* **445**, 65-69 (2007).
19      Takesue, H. *et al.* Quantum key distribution over a 40-dB channel loss using superconducting single-photon detectors. *Nat Photonics* **1**, 343-348 (2007).
20      Ghioni, M., Gulinatti, A., Rech, I., Zappa, F. & Cova, S. Progress in silicon single-photon avalanche diodes. *Ieee J Sel Top Quant* **13**, 852-862 (2007).
21      Eisaman, M. D., Fan, J., Migdall, A. & Polyakov, S. V. Invited Review Article: Single-photon sources and detectors. *Rev Sci Instrum* **82**, 071101 (2011).
22      Itzler, M. A. *et al.* Advances in InGaAsP-based avalanche diode single photon detectors. *J Mod Optic* **58**, 174-200 (2011).

23  Cabrera, B. *et al.* Detection of single infrared, optical, and ultraviolet photons using superconducting transition edge sensors. *Appl Phys Lett* **73**, 735-737 (1998).
24  Hu, X. L., Holzwarth, C. W., Masciarelli, D., Dauler, E. A. & Berggren, K. K. Efficiently Coupling Light to Superconducting Nanowire Single-Photon Detectors. *Ieee T Appl Supercon* **19**, 336-340 (2009).
25  Rosfjord, K. M. *et al.* Nanowire Single-photon detector with an integrated optical cavity and anti- reflection coating. *Opt Express* **14**, 527-534 (2006).
26  Marsili, F. *et al.* Single-Photon Detectors Based on Ultranarrow Superconducting Nanowires. *Nano Lett.* **11**, 2048-2053 (2011).
27  Verevkin, A. *et al.* Ultrafast superconducting single-photon detectors for near-infrared-wavelength quantum communications. *J Mod Optic* **51**, 1447-1458, (2004).
28  Kerman, A. J. *et al.* Kinetic-inductance-limited reset time of superconducting nanowire photon counters. *Appl Phys Lett* **88**, 111116 (2006).
29  Gerrits, T. *et al.* On-chip, photon-number-resolving, telecommunication-band detectors for scalable photonic information processing. *Physical Review A* **84**, 060301 (2011).
30  Sprengers, J. P. *et al.* Waveguide superconducting single-photon detectors for integrated quantum photonic circuits. *Applied Physics Letters* **99,** 181110 (2011).